\documentclass[12pt]{iopart}
\usepackage{latexsym}
\usepackage{graphicx}
\usepackage{amssymb}
\begin{document}

\title{
Transverse-field $XY$ spin chain with the competing
long-range interactions:
Multi-criticality around the $XX$-symmetric point
}

\author{Yoshihiro Nishiyama}

\address{Department of Physics, Faculty of Science,
Okayama University, Okayama 700-8530, Japan}
\vspace{10pt}

\begin{abstract}

The transverse-field $XY$ spin chain
with 
competing antiferromagnetic
long-range interactions,
$J_r \propto 1/r^\alpha$
($r$: distance between spins),
and the exponent $\alpha$
was investigated numerically.
The main concern is to clarify 
the character of
the transverse-field-driven phase transition
for the small-$\alpha$ regime
around the $XX$-symmetric point, 
$\eta=0$ ($\eta$: $XY$-anisotropy parameter).
To cope with the negative-sign problem,
we employed the exact-diagonalization method,
which enables us to evaluate the fidelity susceptibility $\chi_F$.
Because the fidelity susceptibility does not 
assume any order parameter after a phase transition,
it detects the multi-criticality
around the $XX$-symmetric point in a systematic manner.
As a preliminary survey, with $\eta=0.5$ and 
$\alpha=2$ fixed, we made the scaling analysis of $\chi_F$.
The scaling behavior of $\chi_F$ shows that the transverse-field-driven phase transition
belongs to the 2D-Ising universality class.
Thereby, with the properly crossover-scaled $\eta$,
the $\chi_F$ data is cast into the crossover scaling formula
for the small-$\alpha$ regime.
The multi-critical exponents
fed into the scaling formula are argued
through referring to related studies.

\end{abstract}

%
%
%
%
%

\section{\label{section1}Introduction}

The $d$-dimensional O$(N)$-symmetric quantum spin model with the 
ferromagnetic
long-range interactions,
$J_r \propto -1/r^{d+\sigma}$ 
($r$: distance between spins), 
and the exponent $\sigma$
has been investigated extensively so far
\cite{Dutta01,Campana10,Gong16,Defenu17,Maghrebi17,Frerot17,Roy18,Puebla19,Adelhardt20,Giachetti22,Monthus15}; see 
Ref. \cite{Defenu21} for a review.
Intriguingly,
the criticality of
the order-disorder phase transition
corresponds to that of the
$D_{\rm eff}$-dimensional
classical counterpart with the
short-range interactions,
and the effective dimensionality $D_{\rm eff}$ is given 
by the expression, $D_{\rm eff}=1+\frac{2d}{\sigma}$, for $\frac{2d}{3} \le \sigma \le 2$ \cite{Defenu17,Defenu21};
the $N \ge 2$ magnet has some exceptional cases because of the enhanced fluctuations in low dimensions \cite{Defenu17}.
In fairness, it has to be mentioned that
the idea of $D_{\rm eff}$ has been developed 
in the course of the studies of the classical long-range spin models
\cite{Fisher72,Sak73,Gori17,Angelini14,Joyce66}.
As a peculiarity of the quantum long-range magnet,
the correlation length along the real-space direction $\xi$
differs significantly from that of the
imaginary-time direction $\xi_\tau$,
and the 
dynamical critical exponent $z(=\sigma/2)$ 
characterizes the anisotropy $\xi_{\tau}\sim \xi^z$
between these directions quantitatively
\cite{Defenu17,Defenu21}.
Rather technically,
the $\xi$-$\xi_\tau$ anisotropy brings about complications as to the finite-size-scaling analyses.
Apart from the negative sign problem,
the exact diagonalization method has an advantage
in that one does not have to care about the scaling behavior of the imaginary-time
sector,
because the ratio $\xi_\tau/\beta \to 0$
($\beta$: inverse temperature) vanishes 
in the scaling form for the ground-state observables.
Namely, one is able to concentrate only on the scaling behavior of 
the real-space sector,
and a significant simplification is attained even for a non-trivial value of $z \ne 1$.
Hence,
the exponent $z$ can be fixed by the subsequent survey.

Meanwhile,
the transverse-field $XY$ chain with competing antiferromagnetic 
long-range interactions was studied with the series expansion method
\cite{Adelhardt20};
notable features of the Heisenberg and Kitaev magnets are emphasized
in Ref. \cite{Birnkammer22,Diessel22,Vodola16}.
The Ising case has been studied rather extensively \cite{Koffel12,Sun17,Fey16,Koziol21}.
To be specific,
the Hamiltonian for the transverse-field $XY$ 
chain with competing antiferromagnetic long-range interactions is given by 
\begin{equation}
\label{Hamiltonian}
{\cal H}=\sum_{i \ne j}J_{ij}
((1+\eta)S^x_iS^x_j+(1-\eta)S^y_iS^y_j)
-\gamma \sum_{i=1}^N S^z_i.
\end{equation}
Here, the quantum spin-$S=1/2$ operator
${\bf S}_i$ is placed at each site, $i=1,2,\dots,N$.
The summation $\sum_{i\ne j}$ runs over all possible pairs, $i$ and $j$,
and the symbol $J_{ij}$ denotes the corresponding coupling constant.
The variables, $\eta$ and $\gamma$, are the
$XY$ anisotropy and the transverse field, respectively,
and hence, the above-mentioned Ising case \cite{Koffel12,Sun17,Fey16,Koziol21}
corresponds to $\eta=1$.
We imposed periodic boundary conditions,
and the coupling constants take the expression
\begin{equation}
\label{Haldane-Shastry_coupling}
	J_{ij}=  \frac{\sin(\pi /N)^\alpha}{\sin(\pi|i-j|/N)^\alpha}
	,
\end{equation}
as in the
Haldane-Shastry model \cite{Haldane88,Shastry88}
at
$\alpha=2$.
The numerator of this expression (\ref{Haldane-Shastry_coupling})
is meant to fix the energy scale,
because the aim of this paper is to explore the 
criticality including the mean-field one rather than evaluating the transition point $\gamma_c$.
In the ferromagnetic case \cite{Nishiyama19},
even
for rather restricted system sizes $N \le 22$,
such a sinusoidal form of $J_{ij}$ yields a critical exponent
down to $\sigma =0.8$ bounded by error margins,
and the anticipated mean-field behavior for the critical amplitude ratio
was reproduced successfully.
Here, treating even larger system sizes $N \le 32$, we investigate  the 
critical behaviors
including the mean-field type.
For instance, it was shown that
the universality class is retained down to $\alpha \to 0$
owing to the energy-scale fixing
in the ferromagnetic case \cite{Homrighausen17}.
Afterward,
a comparison with the purely-algebraically-decaying case shall be made nonetheless.

\begin{figure}
\includegraphics[width=13cm]{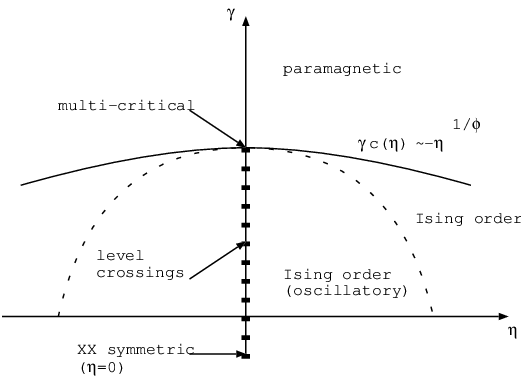}%
\caption{\label{figure1}
A schematic phase diagram for
the transverse-field $XY$ chain 
with competing antiferromagnetic long-range interactions 
(\ref{Hamiltonian})
is shown \cite{Adelhardt20};
here, we also made use of the rigorous information
as for the short-range $XY$ chain 
($\alpha \to \infty$) \cite{Radgohar18,Dutta10}.
For large (small) transverse field, $\gamma>(<)\gamma_c(\eta)$,
the paramagnetic (ordered) phase extends irrespective of the $XY$ anisotropy $\eta$.
The phase boundary $\gamma_c(\eta)$ separates these phases,
and
the crossover exponent $\phi$ 
(\ref{crossover_exponent})
\cite{Riedel69,Pfeuty74}
characterizes the multi-criticality 
(end-point singularity),
$\gamma_c(\eta) \sim -|\eta|^{1/\phi}$,
of the phase boundary
around the $XX$-symmetric point $\eta = 0$.
Along the ordinate axis
$\eta=0$, the level crossings take place
successively
\cite{Rams11,Mukherjee11}, causing
a severe finite-size artifact.
The oscillatory phase extends around the ordinate axis
\cite{Radgohar18,Dutta10}.
The multi-criticality around $\eta=0$ is our concern
particularly for the small decay rate $\alpha$.
}
\end{figure}

A schematic phase diagram for the transverse-field $XY$ chain 
with competing antiferromagnetic long-range interactions
(\ref{Hamiltonian})
is shown in Fig. \ref{figure1} \cite{Adelhardt20}; 
here, we also made use of the rigorous information as for the short-range $XY$ chain 
\cite{Radgohar18,Dutta10}.
For large  (small) transverse fields, $\gamma>(<)\gamma_c(\eta)$, the paramagnetic (ordered)
phase extends irrespective of the $XY$-anisotropy parameter $\eta$.
The phase boundary $\gamma_c(\eta)$ separates these phases. 
The crossover exponent $\phi$ characterizes the
end-point singularity
\cite{Riedel69,Pfeuty74}
\begin{equation}
\label{crossover_exponent}
\gamma_c(\eta)-\gamma_c(0) \sim -\eta^{1/\phi}
,
\end{equation}
of the phase boundary
$\gamma_c(\eta)$
terminating
at the multi-critical point, $\eta=0$.
Along the ordinate axis, $\eta=0$, the level crossings 
take place successively 
\cite{Rams11,Mukherjee11}
due to the
O$(2)$
symmetry around the quantization axis,
giving rise to a severe finite-size artifact.
In fact,
around the ordinate axis,
the correlation function shows an
oscillatory behavior
\cite{Radgohar18,Dutta10}.
As depicted in Fig. \ref{figure1},
a variety of boundaries meet 
at
the multi-critical point
$(0,\gamma_c(0))$.

\begin{figure}
\includegraphics[width=13cm]{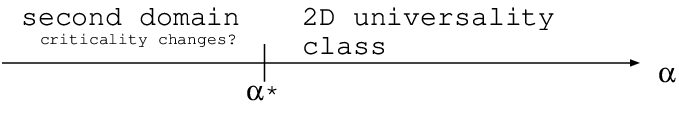}%
\caption{\label{figure2}
A criticality diagram is shown.
The (multi) criticality of the phase boundary
$\gamma_c(\eta)$
(see Fig. \ref{figure1}) may depend on the algebraic decay rate $\alpha$
of the long-range interactions.
Particularly,
at the $XX$-symmetric point, $\eta=0$, the multi-criticality would be sensitive
to $\alpha$,
because
the $XX$ order develops only marginally for the short-range magnet.
Namely, 
for $\alpha>\alpha^*$,
	the $2$D universality class is realized,
whereas 
for sufficiently small $\alpha<\alpha^*$, 
the ``second domain''
	\cite{Adelhardt20}
	may set in.
In the Ising case, $\eta=1$, the 
exitance of such a threshold $\alpha^*$ 
remains controversial \cite{Adelhardt20,Koffel12,Sun17}.  }
\end{figure}

Then, there arises a problem how the decay rate $\alpha$
affects  
the universality class at $\gamma=\gamma_c(\eta)$.
Actually, at the $XX$-symmetric (multi-critical) point,
the singularity may be sensitive to the long-range interactions,
because the $XX$ order develops only marginally for the short-range magnet, $\alpha\to\infty$.
A schematic universality-class diagram
is presented in Fig. \ref{figure2},
where 
a threshold $\alpha^*$
separating
the
$2$D-universality-class regime ($\alpha >\alpha^*$) and the ``second domain''
\cite{Adelhardt20}
($\alpha<\alpha^*$) is assumed.
In the Ising case, $\eta=1$,
the threshold was estimated as
$\alpha^*=2.25$ \cite{Koffel12},
whereas 
no particular signature for $\alpha^*$ could be detected 
\cite{Sun17}
even for the easy-axis case $\eta\ne 0$
\cite{Adelhardt20};
the criticality of the branch $\gamma_c(\eta)$ for $\eta \ne 0$
was studied 
in Ref. \cite{Adelhardt20},
where the parameter $\eta $ is fixed to
a number of fractional values,
$\eta=\frac{1}{3}$, $\frac{1}{2}$,\dots.
The aim of this paper is to shed light on this problem,
devoting ourselves to the
multi-critical point, $\eta=0$, rather than the easy-axis case, $\eta \ne 1$.
Moreover,
to avoid the complications due to the oscillatory behavior
around $\eta\ne0$
\cite{Radgohar18,Dutta10,
Rams11,Mukherjee11},
we took an indirect route towards the multi-critical point $\eta=0$ through properly scaling
$\eta$, based on
the crossover-scaling theory
\cite{Fisher72,Sak73}.
As a byproduct, we estimate the aforementioned crossover exponent $\phi$
(\ref{crossover_exponent}) as well as the multi-critical exponents.

For that purpose,
we employed the exact diagonalization method to surmount the negative sign problem.
Thereby, taking the advantage in that the exact diagonalization method
provides the ground state $| \gamma \rangle$ for the transverse field $\gamma$
explicitly,
we evaluated the fidelity susceptibility \cite{Quan06,Zanardi06,HQZhou08,Yu09,You11}
\begin{equation}
	\label{fidelity_susceptibility}
\chi_F = - \frac{1}{N} \partial_{\Delta \gamma}^2 F(\gamma,\gamma+\Delta \gamma)|_{\Delta \gamma=0}
.
\end{equation}
Here, the fidelity 
$F(\gamma,\gamma+\Delta \gamma)$ is given by
the overlap
\cite{Uhlmann76,Jozsa94,Peres84,Gorin06}
\begin{equation}
F(\gamma,\gamma+\Delta \gamma) =
|
\langle \gamma | \gamma+\Delta \gamma\rangle |
,
\end{equation}
between the ground states with proximate field strengths,
$\gamma$ and $\gamma+\Delta \gamma$.
The singularity of the fidelity susceptibility (\ref{fidelity_susceptibility})
is stronger than that of the specific heat
\cite{Albuquerque10},
and hence, it detects the signature of the criticality more clearly.
Moreover,
it is sensitive to both $XX$- ($\eta=0$) and Ising- ($\eta \ne 1$) symmetric cases,
because the fidelity susceptibility does not 
assume any order parameter after a phase transition.

The rest of this paper is organized as follows.
In Sec. \ref{section2},
we present the numerical results. 
The (crossover) scaling theory for the fidelity susceptibility is explained
prior to the analysis of numerical data.
In Sec. \ref{section3}, we address the summary and discussions.

\section{\label{section2}Numerical results}

In this section, we present the numerical results for the 
transverse-field $XY$ chain with competing antiferromagnetic long-range interactions 
(\ref{Hamiltonian}).
To begin with, we explain the scaling formula 
\cite{Albuquerque10}
for the fidelity susceptibility
(\ref{fidelity_susceptibility}), 
which sets the basis for the present analyses of criticality.
The fidelity susceptibility obeys the scaling formula
\begin{equation}
	\label{scaling_formula}
\chi_F= N^x f
\left((\gamma-\gamma_c)N^{1/\nu}
\right)
,
\end{equation}
with $\chi_F$'s scaling dimension $x$, critical point $\gamma_c$, 
correlation-length critical exponent $\nu$,
and a certain scaling function $f$;
namely, the correlation length $\xi$ diverges as 
$\xi \sim |\gamma-\gamma_c | ^{-\nu}$ at the critical point 
$\gamma=\gamma_c$.
The scaling dimension $x$
satisfies the scaling relation \cite{Albuquerque10}
\begin{equation}
	\label{scaling_relation}
x=\alpha_s /\nu+z
,
\end{equation}
with the specific-heat critical exponent $\alpha_s$ 
and dynamical critical exponent $z$;
namely, the specific heat $C$ exhibits a singularity  
$C \sim |\gamma-\gamma_c|^{-\alpha_s}$.
Notably,
$\chi_F$'s scaling dimension
$x$ is larger than that of the specific heat $\alpha_s/\nu$,
and hence, the fidelity susceptibility
exhibits a pronounced signature of the criticality.

Owing to periodic boundary conditions,
the numerical diagonalization was performed within the Hilbert-space sub-sector
with zero momentum $k=0$.
Within this sector, both ground and excited states exist,
and the
energy gap between these states plays a role in Sec. \ref{section2_3}.
Hence,
it is significant to get rid of influence from the boundaries;
the DMRG method works more efficiently under open boundary conditions
than periodic ones.

\subsection{\label{section2_1}
Transverse-field-driven criticality for the fixed 
anisotropy parameter
$\eta=0.5$}

As a preliminary survey,
we analyze the transverse-field-driven phase transition 
via the fidelity susceptibility 
$\chi_F$
(\ref{fidelity_susceptibility})
with the fixed anisotropy $\eta=0.5$ and algebraic decay rate $\alpha=2$,
for which an elaborated series-expansion result
\cite{Adelhardt20}
is available.
Putting the hyperscaling relation 
$\alpha_s=2-(d+z)\nu$ 
($d$: spatial dimension)
\cite{Albuquerque10}
and $d=1$
into Eq. (\ref{scaling_relation}),
we arrive at
\begin{equation}
\label{scaling_relation_simplified}
x=2/\nu-1
.
\end{equation}
This expression does not depend on 
the dynamical critical exponent $z$,
and hence, the exponent $z$ can be fixed separately. 

\begin{figure}
\includegraphics[width=13cm]{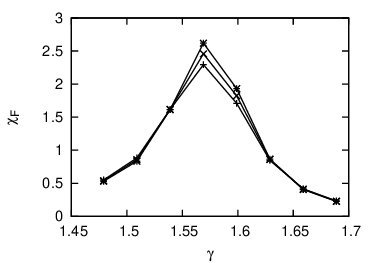}%
\caption{\label{figure3}
The fidelity susceptibility $\chi_F$ (\ref{fidelity_susceptibility}) is shown 
for various transverse field $\gamma$
and system sizes,
($+$) $N=28$, ($\times$) $30$, and ($*$) $32$,
with the fixed 
$XY$ anisotropy $\eta=0.5$ and the algebraic decay rate $\alpha=2$ of the 
	long-range interactions; 
	for the interaction parameters, $(\eta,\alpha)=(0.5,2)$,
	a preceding series-expansion result is available 
	\cite{Adelhardt20}.
	The fidelity susceptibility exhibits a peak
 around $\gamma_c \approx 1.56$, which
indicates an onset of the phase transition between the paramagnetic ($\gamma>\gamma_c$)
and ordered ($\gamma<\gamma_c$) phases.}
\end{figure}

In Fig. \ref{figure3}, we present the
fidelity susceptibility $\chi_F$
for various values of the transverse-field $\gamma$
and system sizes,
($+$) $N=28$
($\times$) $30$,
and
($*$) $32$,
with
$(\eta,\alpha)=(0.5,2)$ fixed.
The fidelity susceptibility exhibits a notable peak
around $\gamma_c \approx 1.56$,
indicating
an onset of the phase transition
separating the paramagnetic ($\gamma>\gamma_c$) and 
ordered ($\gamma< \gamma_c$) phases.

\begin{figure}
\includegraphics[width=13cm]{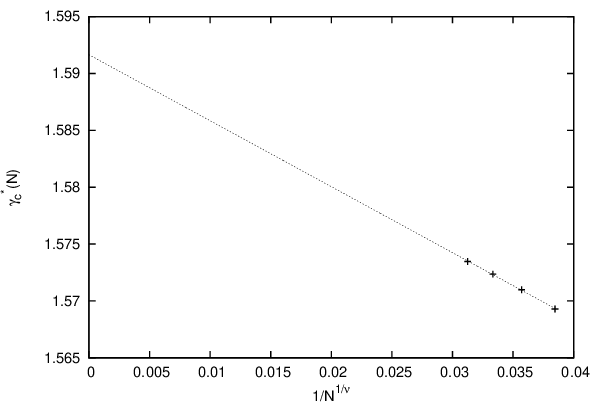}%
\caption{\label{figure4}
The approximate critical point $\gamma_c^*(N)$ 
(\ref{approximate_critical_point})
	is plotted against $1/N^{1/\nu}$ with the correlation-length critical exponent $\nu=1$
	(2D-Ising universality class).
The parameters, $\eta=0.5$ and $\alpha=2$, are the same as those of Fig. \ref{figure3}.
The least-squares fit to those data yields an estimate $\gamma_c=1.5916(4)$ in the thermodynamic limit 
$N\to\infty$.
Possible systematic errors are considered in the text.
To exclude incoherent universality-class data due to the crossover scaling behavior, 
only the $N=26$--$32$ data is taken into account.
}
\end{figure}

In order to estimate the critical point precisely,
in Fig. \ref{figure4},
we present the approximate critical point 
$\gamma_c^*(N)$
against $1/N^{1/\nu}$ 
with the correlation-length critical exponent $\nu=1$
(2D-Ising universality class)
and the fixed $(\eta,\alpha)=(0.5,2)$;
the validity of $\nu=1$ is examined afterward.
In order to exclude incoherent universality-class data due to the
crossover-scaling behavior, only the
$N=26$--$32$ data is taken into account.
The abscissa scale $1/N^{1/\nu}$ comes from the
scaling formula (\ref{scaling_formula}),
which indicates that 
the critical point $\gamma_c$ has the same scaling dimension as that of $1/N^{1/\nu}$,
and hence, the data in Fig. \ref{figure4} should align.
The approximate critical point $\gamma_c^*(N)$ denotes $\chi_F$'s peak position
\begin{equation}
\label{approximate_critical_point}
\partial_\gamma \chi_F|_{\gamma=\gamma_c^*(N)} =0
,
\end{equation}
for each system size $N$.
The least-squares fit to the data in Fig. \ref{figure4}
yields an estimate 
$\gamma_c=1.5916(4)$
in the thermodynamic limit $N\to\infty$.
To appreciate possible systematic errors, 
we made the similar analysis for the system sizes, $N=28$-$32$,
and arrived at an alternative estimate $\gamma_c=1.5909(4)$.
The deviation $\approx 7\cdot10^{-4}$ between them
appears to be comparable with
the least-squares-fit error $\approx 4 \cdot 10^{-4}$,
and hence, the available system sizes 
enter into the scaling regime.
Noticing that both errors are bounded by 
$1\cdot 10^{-3}$, the critical point is estimated as
\begin{equation}
\label{critical_point}
\gamma_c=1.592(1)
.
\end{equation}
In Table \ref{table}, as a reference,
we present the
perturbative continuous unitary transformation
(pCUT) result $\gamma_c|_{\eta=0.5,\alpha=2} = 1.58 $, which is
read off from the data point of
Fig. 4 in Ref. \cite{Adelhardt20}.
In this study \cite{Adelhardt20}, the closure of the energy gap was analyzed
so as
to determine the critical point.
The present exact diagonalization
(ED) result (\ref{critical_point})
is to be compared with
this study \cite{Adelhardt20}.

\begin{table}
\caption{\label{table}
A comparison between the perturbative continuous unitary transformation
(pCUT) result
\cite{Adelhardt20}
and ours is made.
In the former approach \cite{Adelhardt20},
as a quantifier for the criticality,
the energy gap was evaluated,
and the estimate for
	the critical point $\gamma_c|_{\eta=0.5,\alpha=2}=1.58$ is read off from 
the data point of Fig. 4
in Ref. \cite{Adelhardt20}.
Our exact diagonalization (ED) result $\gamma_c=1.592(1)$
[Eq. (\ref{critical_point})]
obtained
via the fidelity susceptibility
(\ref{fidelity_susceptibility})
is to be compared with this pioneering study \cite{Adelhardt20}.
}
\begin{tabular}{@{}llll}
\br
Method & Quantifier & $\gamma_c|_{\eta=0.5,\alpha=2}$ \\
\mr
pCUT \cite{Adelhardt20} & energy gap & $1.58$ \\
	ED (this work) & fidelity susceptibility &  $1.592(1)$ \\
\br
\end{tabular}
\end{table}

\begin{figure}
\includegraphics[width=13cm]{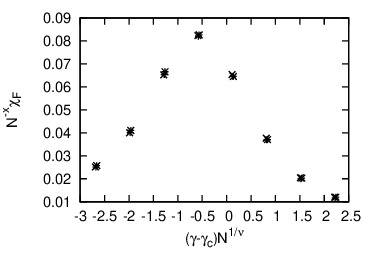}%
\caption{\label{figure5}
	Based on the scaling formula (\ref{scaling_formula}),
$\chi_F$'s scaling plot,
	$(\gamma-\gamma_c)N^{1/\nu}$-$N^{-x} \chi_F$,
is presented for various system sizes,
($+$) $N=28$, ($\times$) $30$, and ($*$) $32$.
The interaction parameters, $\eta=0.5$ and $\alpha=2$,
	are the same as those of Fig. \ref{figure3}.
	The scaling parameters are set to $\gamma_c=1.592$ (\ref{critical_point}),
	$\nu=1$, and $x=1$ (\ref{scaling_dimension}); see text for details.
The result indicates that the criticality belongs to the 2D-Ising universality class
\cite{Adelhardt20,Sun17}.
	}
\end{figure}

We then turn to the scaling analysis of $\chi_F$,
based on the scaling formula (\ref{scaling_formula}).
In Fig. \ref{figure5},
we
present $\chi_F$'s scaling plot,
$(\gamma-\gamma_c)N^{1/\nu}$-$N^{-x} \chi_F$,
for various transverse field $\gamma$ and 
system sizes,
($+$) $N=28$,
($\times$) $30$,
and
($*$) $32$,
with the fixed $\eta=0.5$ and $\alpha=2$.
Here, the scaling parameters are set to
$\gamma_c=1.592$ (\ref{critical_point}),
and
$\nu=1$; namely, we made a proposition that the criticality belongs to
the 2D-Ising universality class
\cite{Adelhardt20,Sun17}.
This exponent $\nu=1$
immediately yields
\begin{equation}
\label{scaling_dimension}
x=1
,
\end{equation}
via Eq.
(\ref{scaling_relation_simplified}).
The scaled data in Fig. \ref{figure5} collapses onto the scaling curve satisfactorily,
validating the 
$\chi_F$-mediated analysis 
as well as the proposition as to the criticality.
We stress
that no
{\it ad hoc} adjustable parameters are undertaken
in the analysis of Fig. \ref{figure5}.

Our observation is not quite accordant with the criticality threshold 
$\alpha^*=2.25$ \cite{Koffel12}.
Rather,
our result shows that the $2$D universality class is retained down to $\alpha=2$,
at least,
in agreement with the claim \cite{Adelhardt20,Sun17}.
Thus,
aiming to
realize $\alpha^*$ as well as the second domain,
we turn
to the analysis of the multi-criticality at $\eta=0$,
where
the $XX$ order would not be so robust against the long-range interactions
as in the easy-axis case $\eta \ne 0$.

Last, we address a remark. 
In the series-expansion analysis of the energy gap \cite{Adelhardt20},
it was claimed that ``(for $\alpha \le 2$)
the critical exponent becomes increasingly challenging to extract.''
In our survey,
rather systematic analysis of criticality could be made
even for $\alpha=2$,
showing that the fidelity susceptibility provides a
reliable indicator even
for such a magnet with competing long-range interactions.

\subsection{\label{section2_2}
Multi-criticality around the $XX$-symmetric point $\eta=0$:
Large-$\alpha$ regime}

We turn to the analysis of the multi-criticality at $\eta=0$ in the 
large-$\alpha$ regime.
Thereby,
we show an evidence that the $2$D multi-criticality is retained 
in this regime,
$\alpha> \alpha^*$.

Because of the level-crossings at the $XX$-symmetric point
$\eta=0$,
the simulation data suffers from irregular finite-size behaviors
\cite{Rams11,Mukherjee11}, as shown in Fig. \ref{figure1}.
Hence,
we took an indirect route to the $XX$-symmetric point
through resorting to the crossover-scaling theory \cite{Riedel69,Pfeuty74}.
That means
incorporating 
yet another scaling parameter $\eta$ accompanying with the crossover exponent $\phi$.
We
extend the aforementioned scaling formula (\ref{scaling_formula})
to
\begin{equation}
\label{crossover_scaling_formula}
\chi_F = N^{\dot{x}}
g\left(\left(\gamma-\gamma_c(\eta)\right) N^{1/\dot{\nu}},
\eta N^{\phi/\dot{\nu}}\right)
,
\end{equation}
\cite{Albuquerque10}
with a certain scaling function $g$.
The multi-critical exponents, $\dot{x}$
and $\dot{\nu}$,
denote 
$\chi_F$'s scaling dimension
and correlation-length critical exponent, respectively,
right at $\eta=0$.
As in Eq. (\ref{scaling_relation}),
The 
exponent
$\dot{x}$ satisfies the
scaling relation
\begin{equation}
\label{crossover_scaling_relation}
\dot{x}=\dot{\alpha}/\dot{\nu} + \dot{z}
,
\end{equation}
with the specific-heat $\dot{\alpha}$
and dynamical $\dot{z}$ critical exponents at $\eta=0$.

Before commencing the scaling analysis of $\chi_F$, 
we fix the multi-critical exponents appearing in Eq. (\ref{crossover_scaling_relation}):
As shown in Fig. \ref{figure2},
we assume that the 
2D-universality-class
phase transition is realized in the large-$\alpha$ regime.
This phase transition, namely,
the transverse-field-driven criticality for the easy-plane 
($\eta=0$) quantum magnet,
has been studied in depth \cite{Zapf14}.
According to this study,
the critical exponents
are given by
$\dot{\alpha}=1/2$,
$\dot{\nu}=1/2$
and 
$\dot{z}=2$.
Hence, from Eq. (\ref{crossover_scaling_relation}),
we obtain
\begin{equation}
\label{crossover_scaling_dimension}
\dot{x}=3
.
\end{equation}
Additionally,
the crossover exponent 
\begin{equation}
\label{crossover_exponent_value}
\phi=1/2
,
\end{equation}
is reported in Ref. \cite{Hofstetter96} 
for the short-range transverse-field  $XY$ spin chain.

\begin{figure}
\includegraphics[width=13cm]{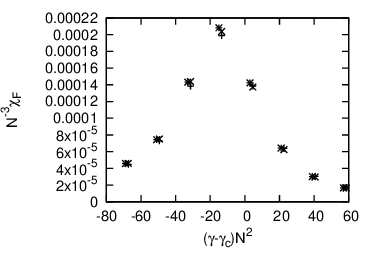}%
\caption{\label{figure6}
	Based on the crossover-scaling formula (\ref{crossover_scaling_formula}),
$\chi_F$'s crossover-scaling plot,
$(\gamma-\gamma_c(\eta))N^{1/\dot{\nu}}$-$N^{-\dot{x}}\chi_F$,
is presented for various system sizes,
($+$) $N=28$, ($\times$) $30$, and ($*$) $32$, with $\alpha=1.5$ fixed.
Here, the second argument of the crossover-scaling formula is fixed to 
$\eta N^{\phi/\dot{\nu} }=12.8$ with the crossover exponent
$\phi=1/2$
	(\ref{crossover_exponent_value}).
The multi-critical exponents are set to $\dot{\nu}=1/2$, and $\dot{x}=3$
	(\ref{crossover_scaling_dimension}),
	and the critical point $\gamma_c(\eta)$ was determined
	via the same scheme as that of Sec. \ref{section2_1};
see text for details.
The result indicates that the multi-criticality at $\eta=0$
	is identical to that of the
	$2$D
	easy-plane ($\eta=0$) magnet in the large-$\alpha(>\alpha^*)$ regime.}
\end{figure}

In Fig. \ref{figure6}, we present the crossover scaling plot, 
$(\gamma-\gamma_c(\eta))N^{1/\dot{\nu}}$-$N^{-\dot{x}}\chi_F$,
for various 
system sizes,
($+$) $N=28$,
($\times$) $30$,
and
($*$) $32$,
with the fixed $\alpha=1.5$.
Here,
the scaling parameters are set to the above values,
$\dot{\nu}=1/2$,
and
$\dot{x}=3$ (\ref{crossover_scaling_dimension}),
and the critical point $\gamma_c(\eta)$ was determined
via the same scheme as that of Sec. \ref{section2_1}.
The second argument of the crossover-scaling formula 
(\ref{crossover_scaling_formula})
is fixed to $\eta N^{\phi/\dot{\nu}}(=0.4\cdot 32^{\phi/\dot{\nu}})=12.8$
with the crossover exponent $\phi=1/2$ (\ref{crossover_exponent_value});
such a constraint $\eta N^{\phi/\dot{\nu}}= C$
ensures that the parameter $\eta$
approaches to $\eta\to 0$ ($N\to\infty$)
in a systematic manner so that
the multi-criticality is captured properly.
The scaled data in Fig. \ref{figure6}
collapses onto a scaling curve satisfactorily,
showing that the multi-criticality belongs to the 
same universality class as that of the $2$D easy-plane 
($\eta=0$) magnet \cite{Zapf14}.
As mentioned in the introduction, such a feature has been reported 
for the
Ising ($\eta=1$) \cite{Sun17} 
as well as the easy-axis 
($\eta\ne0$) \cite{Adelhardt20} magnets.
We stress that the scaling exponents,
$\dot{\nu}=1/2$,
$\dot{x}=3$ (\ref{crossover_scaling_dimension}),
and
$\phi=1/2$ (\ref{crossover_exponent_value}),
are all fixed
prior to the scaling analysis.
Therefore,
no adjustable parameters are undertaken in 
the present scaling analysis,
Fig. \ref{figure6}.

We address a remark.
The underlying physics behind the crossover scaling plot shown in Fig. \ref{figure6}
differs from that of Fig. \ref{figure5} with the fixed $\eta=0.5$.
Actually. the scaling dimension of the former $\dot{x}=3$ 
(\ref{crossover_scaling_dimension})
is substantially larger than the latter
$x=1$ (\ref{scaling_dimension}),
indicating that
the data collapse of the each scaling plot is by no means accidental.

\subsection{\label{section2_3}
Multi-criticality around the $XX$-symmetric point $\eta=0$:
Small-$\alpha$ regime}

In this section, we investigate the small-$\alpha$ regime.
In order to analyze the simulation results as in Sec. \ref{section2_2}, we need to postulate a set of exponents
appearing in
the crossover-scaling formula (\ref{crossover_scaling_formula});
the reasoning behind these exponents will be argued
afterwards
in the context of the finite-size-scaling theory above the upper critical dimension.
For that purpose,
the correlation-length and dynamical exponents are fixed to
\begin{equation}
\label{multi-critical_correlation_length_exponent2}
\dot{\nu}=1/\alpha
,   
\end{equation}
and $z=\alpha/2$, respectively.
As mentioned in Sec. \ref{section2_2}, according to the study of the transverse-field-driven phase transition
for the easy-plane magnet \cite{Zapf14},
the specific-heat and dynamical critical exponents are given by
$\dot{\alpha}=1/2$, and 
\begin{equation}
\label{multi-critical_dynamical_exponent2}
\dot{z}(=2z)=\alpha
,
\end{equation}
respectively.
Hence, feeding these exponents into
Eq. (\ref{crossover_scaling_relation}),
we obtain $\chi_F$'s scaling dimension
\begin{equation}
\label{crossover_scaling_dimension2}
\dot{x}=3\alpha/2 ,
\end{equation}
at $\eta=0$.

\begin{figure}
\includegraphics[width=13cm]{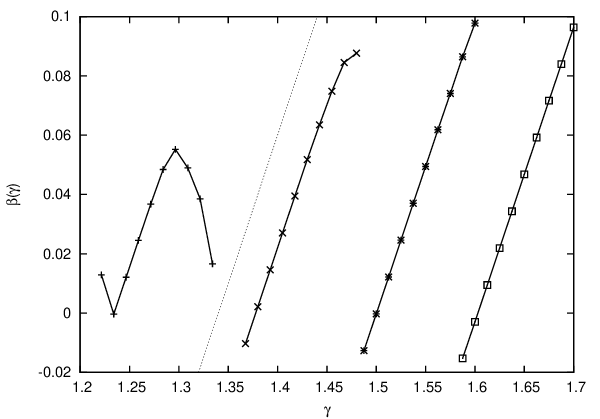}%
\caption{\label{figure7}
The $\beta$ function (\ref{beta_function}) is plotted for various $\gamma$,
and 
	($+$) $\alpha=0.6$, ($\times$) $1$, ($*$) $1.4$, and ($\Box$) $1.8$, with $\eta=0.2$ and $N=22$.
	The dynamical critical exponent $z=1$ (2D-Ising universality class) was put into the formula (\ref{beta_function}) tentatively.
	The zero points of $\beta$ locate the critical points as
	$\gamma_c=1.23$,
	$1.38$,
	$1.5$, and
	$1.6$, respectively,
	and the slope $1/\nu$ at each $\gamma_c$
	seems to obey the
	the 2D-Ising-universality-class behavior, {\it i.e.}, $1/\nu=1$, as indicated by
	the dotted line.
	Therefore, the 2D-Ising universality class is retained down to $\alpha=0.6$,
	at least,
in accordance with Ref. \cite{Adelhardt20,Sun17}.
	Notable signature from the crossover-criticality is seen at $\alpha=0.6$,
	which is identified as a crossover-criticality different from that of Sec. \ref{section2_2}
	by the subsequent analyses.
}
\end{figure}

As a preliminary survey,
we evaluated
the $\beta$ function \cite{Roomany80}.
The beta function $\beta(\gamma)$ describes coupling-constant-$\gamma$'s flow through the renormalization
group
such that the renormalized $\gamma$ increases (decreases) for $\beta>(<)0$.
The scale-invariant point $\beta(\gamma)=0$ locates the onset of the phase transition, 
$\gamma=\gamma_c$, and
the slope of the $\beta$ function indicates
the inverse correlation-length critical exponent 
$1/\nu$.
Hence, 
we are able to observe
how the multi-criticality comes out from 
the overall (crossover) behavior of the $\beta$ function.
To be specific,
the Roomany-Wyld approximant of the $\beta$ function
\cite{Roomany80}
is given by the expression
\begin{equation}
	\label{beta_function}
\beta(\gamma)=\frac{\dot{z}+\ln(\Delta E(N+2)/\Delta E(N))/\ln((N+2)/N)}
{\sqrt{
\frac{
\partial_\gamma \Delta E(N+2) \partial_\gamma \Delta E(N)
}{
\Delta E(N+2)\Delta E(N)
}
}
}
,
\end{equation}
with the energy gap $\Delta E(N)$ for each $N$,
which is readily evaluated by means of
the exact diagonalization method.
In Fig. \ref{figure7},
we present the $\beta$ function for various $\gamma$,
and
	($+$) $\alpha=0.6$, ($\times$) $1$, ($*$) $1.4$, and ($\Box$) $1.8$, with $\eta=0.2$ and $N=22$.
	Here, the dynamical critical exponent $z=1$ (2D-Ising universality class) was put into the 
	formula (\ref{beta_function}) tentatively,
	and the $\gamma<\gamma_c$ data 
	is eliminated because of the irregularity 
	due to the oscillatory-Ising-ordered phase.
As mentioned above,
from the condition $\beta(\gamma_c)=0$,
the critical points are estimated as
	$\gamma_c \approx 1.23$,
	$1.38$,
	$1.5$, and
	$1.6$, respectively.
The slope at each $\gamma=\gamma_c$ seems to obey the
	2D-Ising-universality-class  behavior, {\it i.e.},
	$1/\nu=1$, as indicated by the dotted line.
	Therefore, for $\eta(=0.2)\ne 0$,
	the 2D-Ising universality class appears to be retained
	down to $\alpha=0.6$, at least, in accordance with
Ref. \cite{Adelhardt20,Sun17}.
More specifically, the width of the $1/\nu=1$-linear regime,
namely, the scaling regime, shrinks,
	as $\alpha$ decreases, particularly for $\alpha=0.6$.
	We identified this feature at $\alpha=0.6$ as the influence from
	the multi-critical point (crossover phenomenon),
	which turns out to be
	different from that of Sec. \ref{section2_2} in the subsequent survey.

\begin{figure}
\includegraphics[width=13cm]{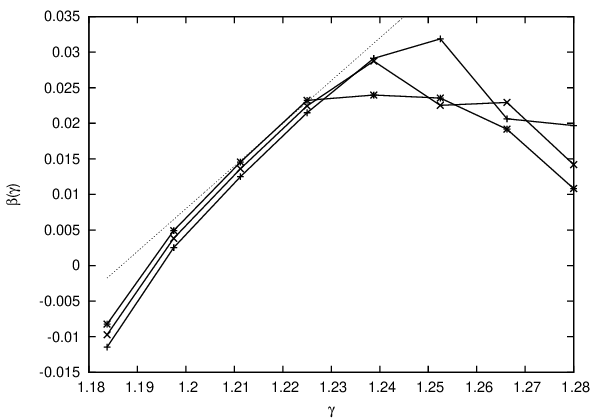}%
\caption{\label{figure8}
The $\beta$ function (\ref{beta_function}) is plotted for various $\gamma$,
and 
($+$) $N=26$, ($\times$) $28$, and ($*$) $30$, with $\eta=0.4$ and $\alpha =0.6$ fixed.
As a dotted line,
the slope $1/\dot{\nu}=\alpha$ (\ref{multi-critical_correlation_length_exponent2}) is shown.
The multi-criticality emerges
in the regime $1.2<\gamma<1.24$.	
Because the postulation $\dot{z}=\alpha$ (\ref{multi-critical_dynamical_exponent2})
is validated in the limit $\eta\to0$,
such a feature should be interpreted a precursor for the multi-criticality.
}
\end{figure}

We then made the
proposition $\dot{z}=\alpha$ 
(\ref{multi-critical_dynamical_exponent2}) 
that the criticality is described by the 
aforementioned multi-critical exponents.
Because
this proposition is validated in the limit $\eta \to 0$,
the $\beta$ function may 
capture 
how the crossover from the multi-critical to the ordinary-2D
universality classes
takes place.
In Fig. \ref{figure8},
we present the $\beta$ function for various $\gamma$,
and system sizes,
($+$) $N=26$
($\times$) $28$,
and
($*$) $30$,
with the fixed $\eta=0.4$ and $\alpha=0.6$.
As a dotted line, we show the slope of $1/\dot{\nu}=\alpha=0.6$
(\ref{multi-critical_correlation_length_exponent2}).
We see that the $\beta$ function detects the aforementioned
multi-criticality
$1/\dot{\nu}=\alpha$ 
in the regime $1.2<\gamma<1.24$; 
we stress that the slope differs significantly from $1/\nu=1$ (2D-Ising universality class).
Rather intriguingly,
in proximity to
the critical point $\gamma=\gamma_c(\approx 1.2)$, 
the slope of the $\beta$ function deviates from the multi-critical value
$\alpha=0.6$ (dotted line), 
indicating that the ordinary 2D criticality emerges in close vicinity of the critical point
$\gamma=\gamma_c$.
Namely, in the first place, the multi-critical fluctuations emerge as a precursor for the phase transition,
and eventually, the 2D-Ising universality class dominates the singularity in proximity to the critical point.
As for $\eta \to 0$, 
this
precursor regime may coincide with the critical point $\gamma=\gamma_c$
so that the second domain sets in.

\begin{figure}
\includegraphics[width=13cm]{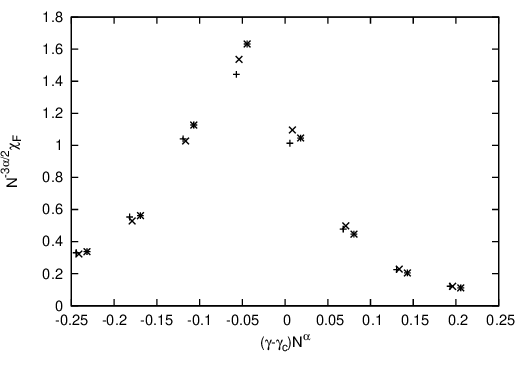}%
\caption{\label{figure9}
	Based on the crossover-scaling formula (\ref{crossover_scaling_formula}),
	$\chi_F$'s crossover-scaling plot,
$(\gamma-\gamma_c(\eta))N^{1/\dot{\nu}}$-$N^{-\dot{x}}\chi_F$,
is presented for various system sizes,
($+$) $N=28$, ($\times$) $30$, and ($*$) $32$, with $\alpha=0.6$ fixed.
Here, the second argument of the crossover-scaling formula 
	(\ref{crossover_scaling_formula}) is fixed to $\eta N^{\phi/\dot{\nu}}=1.131$
	with the crossover exponent $\phi=1/2$
	(\ref{crossover_exponent_value}),
and the multi-critical exponents are set to 
$\dot{\nu}=1/\alpha$ (\ref{multi-critical_correlation_length_exponent2}),
and
$\dot{x}=3\alpha/2$ (\ref{crossover_scaling_dimension2}).
	Except for the crossover exponent $\phi$,
	the other exponents differ from those of Fig. \ref{figure6},
	indicating that the multi-criticality enters into the second domain, as 
	shown in Fig. \ref{figure2}.
	}
\end{figure}

We then analyze the multi-criticality via the fidelity susceptibility, based on the
crossover-scaling formula (\ref{crossover_scaling_formula})
with the
above-mentioned multi-critical exponents, 
 $\dot{\nu}$ (\ref{multi-critical_correlation_length_exponent2}),
and $\dot{x}$ (\ref{crossover_scaling_dimension2}).
In Fig. \ref{figure9}, we present the crossover-scaling plot,
$(\gamma-\gamma_c(\eta))N^{1/\dot{\nu}}$-$N^{-\dot{x}} \chi_F$,
for various 
system sizes,
($+$) $N=28$,
($\times$) $30$,
and
($*$) $32$,
with the decay rate $\alpha=0.6$.
As mentioned above,
the scaling parameters are set to 
$\dot{\nu}=1/\alpha$ (\ref{multi-critical_correlation_length_exponent2}),
and
$\dot{x}=3\alpha/2$ (\ref{crossover_scaling_dimension2}).
The second argument of the crossover-scaling formula (\ref{crossover_scaling_formula})
is fixed to $\eta N^{\phi/{\dot{\nu}}}(=0.4\cdot32^{\phi/\dot{\nu}}) \approx 1.131$
with the same crossover exponent $\phi=1/2$ as that of Fig. \ref{figure6}.
the crossover-scaled data seems to collapse onto the scaling curve satisfactorily.
We stress that the undertaken exponents, $\dot{\nu}=1/\alpha$ and $\dot{x}=3\alpha/2$,
differ from those of Fig. \ref{figure6}, 
{\it i.e.},
$\dot{\nu}=1/2$ and 
$\dot{x}=3$, significantly.
Therefore, the result of Fig. \ref{figure9} shows that 
the multi-criticality enters into the second domain for $\alpha=0.6$;
see Fig. \ref{figure2}.

The crossover-scaling formula
(\ref{crossover_scaling_formula}) 
and related preceding studies are reconciled
by the scaling theory above the upper critical dimension
\cite{Langheld22,Jones05}.
Following this theory \cite{Jones05},
we introduce the
length scale
\begin{equation}
	\label{length_l}
l=N^{\alpha/2}
.
\end{equation}
Converting the length scale $N$ into $l$,
the crossover-scaling formula (\ref{crossover_scaling_formula}) 
accompanied with
$\dot{\nu}=1/\alpha$ (\ref{multi-critical_correlation_length_exponent2})
and
$\dot{x}=3\alpha/2$ (\ref{crossover_scaling_dimension2})
is rewritten as
\begin{equation}
	 	\chi_F=l   ^{\dot{x}_{l}}
		g((\gamma-\gamma_c) l^{1/\dot{\nu}_l},\eta l^{\phi/\dot{\nu}_l})
		,
\end{equation}
Here, the critical exponents,
$\dot{x}_l$ and $\dot{\nu}_l$, as well as 
$\dot{z}_l$ defined by
$\xi_\tau (\sim N^{\dot{z}})\sim l^{\dot{z}_l}$ and
Eq. (\ref{multi-critical_dynamical_exponent2})
reduce to
\begin{equation}
	\label{l-mediated_exponents}
	\dot{x}_l=3, \ 
	\dot{\nu}_l=1/2 \
	{\rm and} \ \dot{z}_l=2
	, 
\end{equation}
respectively.
These critical exponents are identical to those of the magnetic-field-induced magnon-condensation transition \cite{Zapf14},
and actually, those exponents have been used in Sec. \ref{section2_2};
detailed microscopic analyses of the magnon dispersion relation as well as its gap closure
for the long-range magnets
were reported in Ref. \cite{Adelhardt20}, where the thermodynamic limit $N\to \infty$
is taken safely.
As in the ordinary magnet,
above the upper critical dimension $\alpha < \alpha^*$,
the exponents at $\alpha=\alpha^*$ continue to be realized
through taking the thermodynamic limit.
Moreover, the threshold 
$\alpha^*=2/3$ 
is also
derived
\cite{Dutta01,Langheld22}
by inserting the above critical exponents into the hyperscaling relation.
Therefore, the scaling theory (\ref{crossover_scaling_formula})
does not contradict these preceding results.

We address a remark.
At the threshold $\alpha=\alpha^*$,  
notorious logarithmic corrections to scaling
\cite{Fey16,Luijten02,Brezin14,Defenu15} appear.
Therefore, the transient behavior around $\alpha=\alpha^*$
was not pursued here in detail.

\section{\label{section3}
Summary and discussions}

The transverse-field $XY$ spin chain with competing antiferromagnetic long-range interactions
(\ref{Hamiltonian})
\cite{Adelhardt20}
was investigated with the exact diagonalization method.
So far,
the Ising case $\eta=1$ has been studied in depth \cite{Koffel12,Sun17,Fey16,Koziol21},
and it has been claimed that
the $2$D criticality is retained down to a considerably small $\alpha$ regime
\cite{Adelhardt20,Sun17}.
In contrast, at the $XX$-symmetric point $\eta=0$,
the magnetism would be
sensitive to $\alpha (< \alpha^*)$, because
the $XX$-order 
develops only marginally 
for the short-range magnet.
In this paper, 
to avoid the level crossings at $\eta=0$,
we took an indirect route towards $\eta=0$,
relying on
the crossover-scaling formula
(\ref{crossover_scaling_formula}).
As a preliminary survey,
with the fixed $\eta=0.5$ and $\alpha=2$,
we investigated
the transverse-field-driven phase transition
via 
the fidelity susceptibility $\chi_F$ (\ref{fidelity_susceptibility}).
The scaled $\chi_F$ data shows that the criticality belongs to the 2D-Ising universality class
\cite{Adelhardt20,Sun17}.
Thereby, with the properly crossover-scaled $\eta$,
the $\chi_F$ data was cast into the crossover scaling formula (\ref{crossover_scaling_formula}).
It turned out that for small $\alpha=0.6$, eventually,
the multi-criticality enters into the second domain, and the criticality is
well described by 
the scaling formula (\ref{crossover_scaling_formula}) accompanied with 
the proposition,
$\dot{\nu}=1/\alpha$ (\ref{multi-critical_correlation_length_exponent2})
and
$\dot{x}=3\alpha/2$ (\ref{crossover_scaling_dimension2}).
These exponents reduce to the feasible expressions,
	$\dot{\nu}_l=1/2$ and
	$\dot{z}_l=2$, as shown in Eq.  (\ref{l-mediated_exponents}).
The present scheme is designed so as to approach to the multi-critical point indirectly
via the crossover-scaling formalism.
Hence, 
the direct simulation results right at the $XX$-symmetric point,
such as
the 
magnon dispersion relation,
are not available.
In this respect,
it has to be mentioned that the present results are intended to appreciate
the power-law behavior of the fidelity susceptibility
through the track guided by
the crossover-scaling ansatz.

Around the threshold $\alpha^*=2/3$ derived 
by the hyperscaling relation
\cite{Dutta01,Langheld22},
there should appear
notorious
logarithmic corrections to scaling \cite{Fey16,Luijten02,Brezin14,Defenu15}, which prevent us 
from investigating the transient behavior in detail.
It would be tempting to subtract these scaling corrections \cite{Fey16}
so as to look into the leading singularities clearly.
This problem is left for the future study.

\ack

This work was supported by a Grant-in-Aid
for Scientific Research (C)
from Japan Society for the Promotion of Science
(Grant No. 20K03767).



\end{document}